\documentclass[prl,twocolumn,showpacs,showkeys,preprintnumbers,superscriptaddress,amsmath,amssymb]{revtex4-1}
\usepackage{graphicx}% Include figure files
\usepackage{bm}% bold math
\usepackage{amsmath}
\usepackage{times}
\usepackage{amsfonts}

\newcommand{\lamno}{(LaMn$_{3}$)Mn$_{4}$O$_{12}$}
\newcommand{\camno}{(CaMn$_3$)Mn$_4$O$_{12}$}

\begin{document}

\title{Ferroelectricity in the 1 $\mu$C cm$^{-2}$ range induced by canted antiferromagnetism in \lamno}

%% Notice placement of commas and superscripts and use of &
%% in the author list
\author{A. Gauzzi}
\email[Corresponding Author.~E-mail: ]{andrea.gauzzi@upmc.fr}
\affiliation{IMPMC, Sorbonne Universit\'e, CNRS, IRD, MNHN, 4, place Jussieu, 75005 Paris, France}
\author{F. Milton}
\author{V. Pascotto Gastaldo}
\affiliation{Departamento de F\'isica, Universidade Federal de S\~ao Carlos, Rodovia Washington Lu\'is, km 235, S\~ao Carlos - S\~ao Paulo, Brazil}
\author{M. Verseils}
\affiliation{IMPMC, Sorbonne Universit\'e, CNRS, IRD, MNHN, 4, place Jussieu, 75005 Paris, France}
\author{A. Gualdi}
\author{D. Dreifus}
\affiliation{Departamento de F\'isica, Universidade Federal de S\~ao Carlos, Rodovia Washington Lu\'is, km 235, S\~ao Carlos - S\~ao Paulo, Brazil}
\author{Y. Klein}
\affiliation{IMPMC, Sorbonne Universit\'e, CNRS, IRD, MNHN, 4, place Jussieu, 75005 Paris, France}
\author{D. Garcia}
\author{A.J.A de Oliveira}
\affiliation{Departamento de F\'isica, Universidade Federal de S\~ao Carlos, Rodovia Washington Lu\'is, km 235, S\~ao Carlos - S\~ao Paulo, Brazil}
\author{P. Bordet}
\affiliation{Institut Ne\'el, CNRS and Universit\'e Grenoble Alpes, BP 166, 38042 Grenoble, France}
\author{E. Gilioli}
\affiliation{Istituto dei Materiali per Elettronica e Magnetismo-CNR, 
Area delle Scienze, 43100 Parma, Italy}

\date{\today}

\begin{abstract}
Pyroelectric current and magnetoelectric coupling measurements on polycrystalline samples of the quadruple perovskite \lamno~ give evidence of ferroelectricity driven by the antiferromagnetic ordering of the $B$-site Mn$^{3+}$ ions at $T_{N,B}$=78 K with record values of remnant electric polarization up to $P$=0.56 $\mu$C cm$^{-2}$. X-ray diffraction measurements indicates an anomalous behavior of the monoclinic $\beta$ angle at $T_{N,B}$, which suggests that $P$ lies in the $ac$-plane, where the moments are collinear, so we conclude that exchange striction is the mechanism of spin-driven ferroelectricity. Polarization values $\sim$3 $\mu$C cm$^{-2}$ are expected in single crystals, which would open the avenue towards practical multiferroic applications.  
\end{abstract}

\maketitle

In the search for suitable multiferroic materials for applications, there has been an increasing interest in magnetic ferroelectrics, where ferroelectricity is induced by a magnetic order. While this mechanism leads to inherently strong magnetoelectric couplings \cite{eer06}, it turns out that the remnant polarizations hitherto reported in magnetic ferroelectrics are typically $P \sim 0.1 \mu$C cm$^{-2}$ or smaller, with the exception of the perovskitelike compound TbMnO$_3$, where values $\sim 1 \mu$C cm$^{-2}$ have been reached under high pressures \cite{aoy14}. It is believed that such small values reflect the weakness of the spin-orbit interaction in non-collinear spin structures \cite{che07}, where the polarization arises from the antisymmetric exchange striction (inverse Dzyaloshinskii-Moriya interaction) between neighbouring $i$ and $j$ spins, \textit{i.e.} $P \propto \mid \bm{{\rm s}}_i \times \bm{{\rm s}}_j \mid$. The challenge is that much higher values in the 10 $\mu$C cm$^{-2}$ range are required for applications. In order to address this issue, the attention has been drawn to collinear spin structures, where large $P \propto \mid \bm{{\rm s}}_i \cdot \bm{{\rm s}}_j \mid \gtrsim 1 \mu$C cm$^{-2}$ induced by the stronger symmetric exchange striction have been predicted \cite{ser06}. This possibility is supported by the observation of stronger ferroelectricity induced by a collinear $E$-structure in the above $R$MnO$_3$ system ($R$ rare earth). A further indication of the role of the symmetric exchange striction is given by a remarkable enhancement of $P$ up to 0.27 $\mu$C cm$^{-2}$ in single crystalline samples of another manganese oxide \camno~ \cite{zha11,joh12} (CMO) characterized by the \textit{quadruple perovskite} $AA'_3B_4$O$_{12}$ structure, which contains two distinct $A'$ and $B$ Mn sites \cite{mar73}. This finding triggered a number of studies aiming at understanding the origin of this enhancement \cite{lux12,zha13}, which may foster the design of multiferroics with even larger $P$ values. However, in CMO, the polarization is concomitant to a complex interplay of an incommensurate helicoidal spin structure and of a charge and orbital ordering of the Mn$^{3+}$ and Mn$^{4+}$ ions \cite{joh16}. This hampers a comparison between experiment and theory and to single out the distinct contributions of the symmetric and antisymmetric exchange terms to the polarization.

In view of the above, model systems, where the above two contributions to ferroelectricity can be clearly identified and controlled, would be very useful to elucidate the mechanism of $P$ enhancement in magnetic multiferroics. Here, we focus on \lamno \cite{pro09} (LMO), where similar multiferroic properties are expected, for CMO shares with CMO a similar quadruple perovskite $AA'_3B_4$O$_{12}$ structure and similar electronic properties. Our motivation is that, in spite of these similarities, LMO exhibits comparatively simple structural and electronic properties, namely single-valent Mn$^{3+}$ characteristics and a commensurate canted $C$-type antiferromagnetic (AFM) order of the $B$-site Mn$^{3+}$ ions at $T_{N,B}$=78 K. Contrary to CMO, LMO displays neither charge orderings, nor incommensurate structural modulations. Its crystal structure undergoes a cubic $Im\overline{3}$ to monoclinic $I2/m$ distortion at 653 K \cite{oka09} and no further structural phase transitions are observed down to 3.5 K within the experimental resolution of powder neutron diffraction \cite{pro09}. 

The LMO samples were synthesized under high-pressure as described elsewhere \cite{pro09}. The as-prepared samples are cylinders of 4 mm size in both diameter and height. Depending on the exact temperature and pressure conditions, either powders or single crystals are obtained. A previous powder neutron diffraction structural study \cite{pro09} reveals that the samples are 95 \% pure and contain minor LaMnO$_3$ and Mn$_3$O$_4$ impurities. In order to unveil the interplay between magnetic and structural properties, a powder sample was investigated by means of x-ray diffraction as a function of temperature in a kappaCCD diffractometer equipped with a liquid He cryostat. The unit cell parameters were refined using the FullProf package. The multiferroic properties were investigated by means of pyrocurrent, $I_p$, measurements on two 0.2 mm thin disks of sintered LMO powders of the same batch. The electrodes were realized by depositing a $0.1 \mu$m thick Au layer on both sides of the disks by magnetron sputtering. The disks were mounted in a closed-cycle cryogenic system and cooled down to 100 K. At this temperature, a dc electric field in the 10 - 36 kV cm$^{-1}$ range was applied by means of a high dc voltage source for 30 minutes to ensure the full polarization of the sample. The samples were subsequently cooled down to 15 K. At this temperature, the electric field was removed and a short-circuit was applied for 10 minutes to minimize space charge effects. Finally, $I_p$ was measured upon warming-up the samples up to 150 K at variable rates of 2 - 5 K min$^{-1}$. The above study was complemented by field- and temperature-dependent specific-heat, $C(T,H)$, measurements on the same powder samples using the 2$\tau$ relaxation method in a PPMS apparatus. The measurements were carried out in the zero-field cooling mode by ramping the field, $H$, up to 9 T at constant temperature in the 2-60 K range.

\begin{figure}[h]
\centering
\includegraphics[width=\columnwidth]{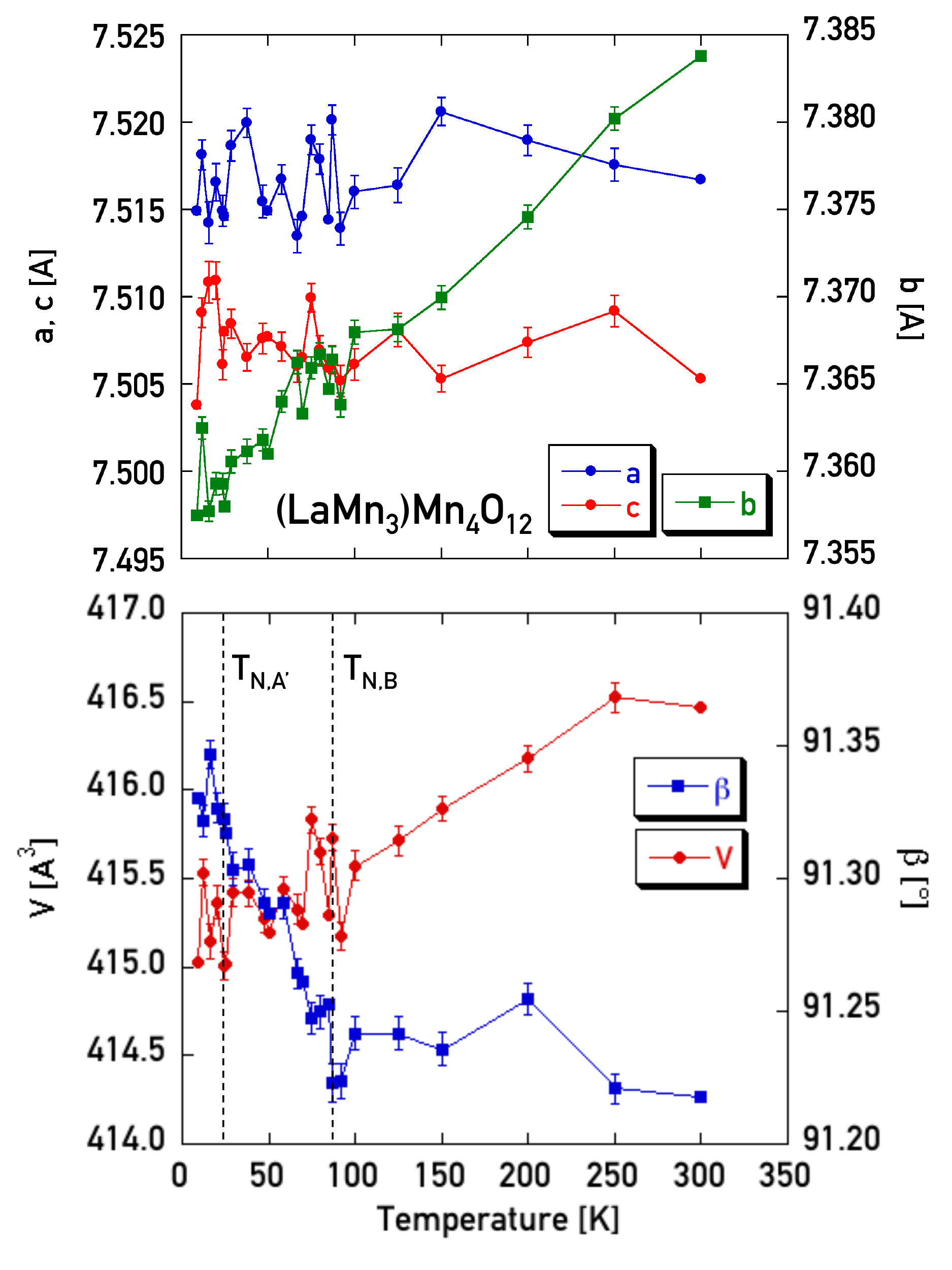}
\caption{\label{fig:cellpar} Temperature dependence of the unit cell parameters and of the unit cell volume, $V$, of the monoclinic crystal structure of \lamno~ powders. Note the sudden increase of the monoclinic angle, $\beta$, at the AFM ordering temperature of the $B$-site Mn$^{3+}$ ions, $T_{N,B}$=78 K, which indicates a large magnetoelastic coupling.}
\end{figure}

\begin{figure}[h]
\centering
\includegraphics[width=\columnwidth]{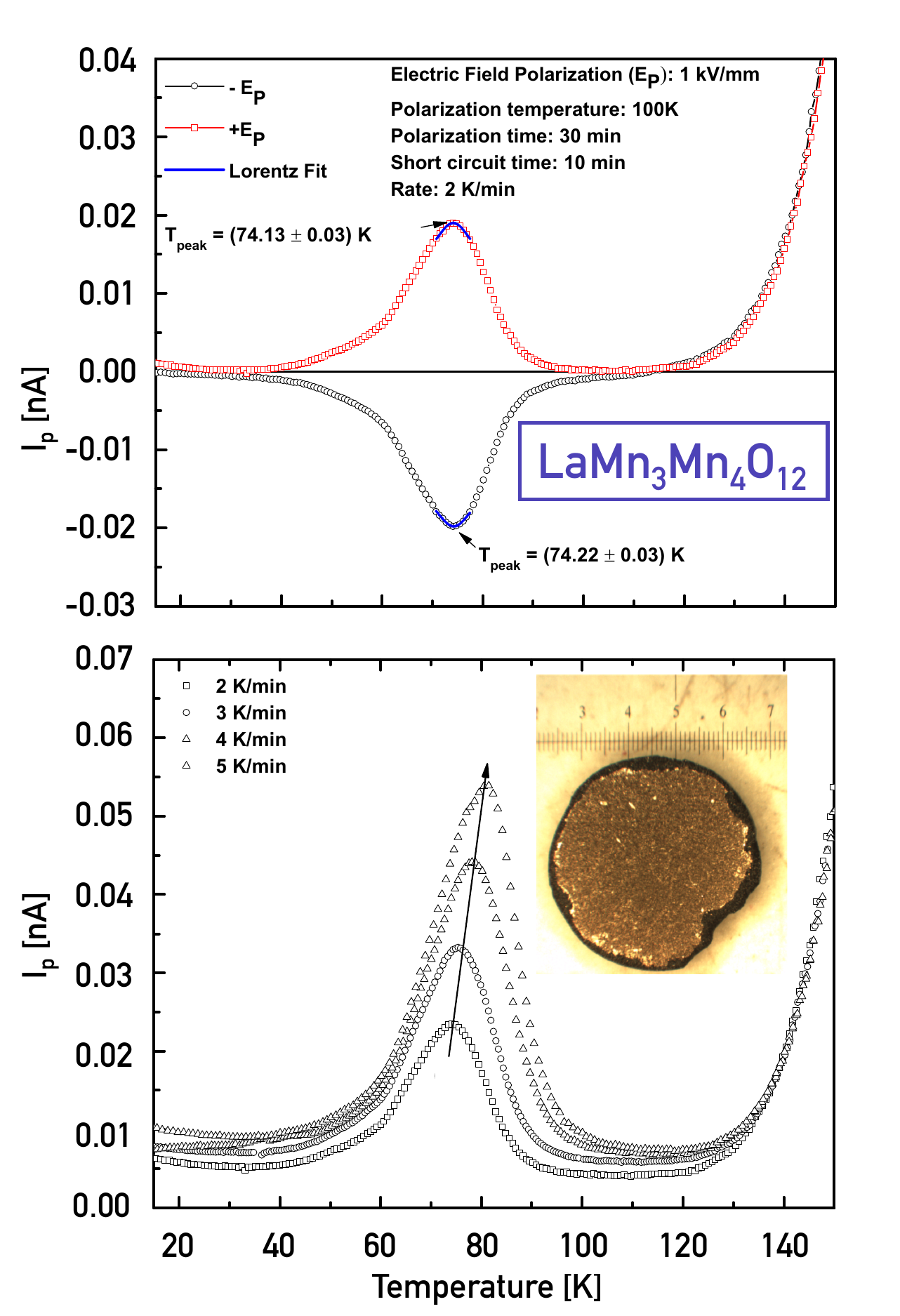}
\caption{\label{fig:pyro} Top panel: Pyrocurrent, $I_p$, curves measured on sample nr. 2 at a heating rate of 2 K/min for a poling field $E_p=\pm10$ kV cm$^{-1}$. Note the $I_p$, peak located near the antiferromagnetic ordering temperature, $T_{N,B}$=78 K. Bottom panel: evolution of the peak with heating rate. The arrow indicates the shift of peak position. The inset shows the 0.2 mm thick disk made of LMO sintered powders covered by a Au layer for metallization.}
\end{figure}

\begin{figure}[h]
\centering
\includegraphics[width=\columnwidth]{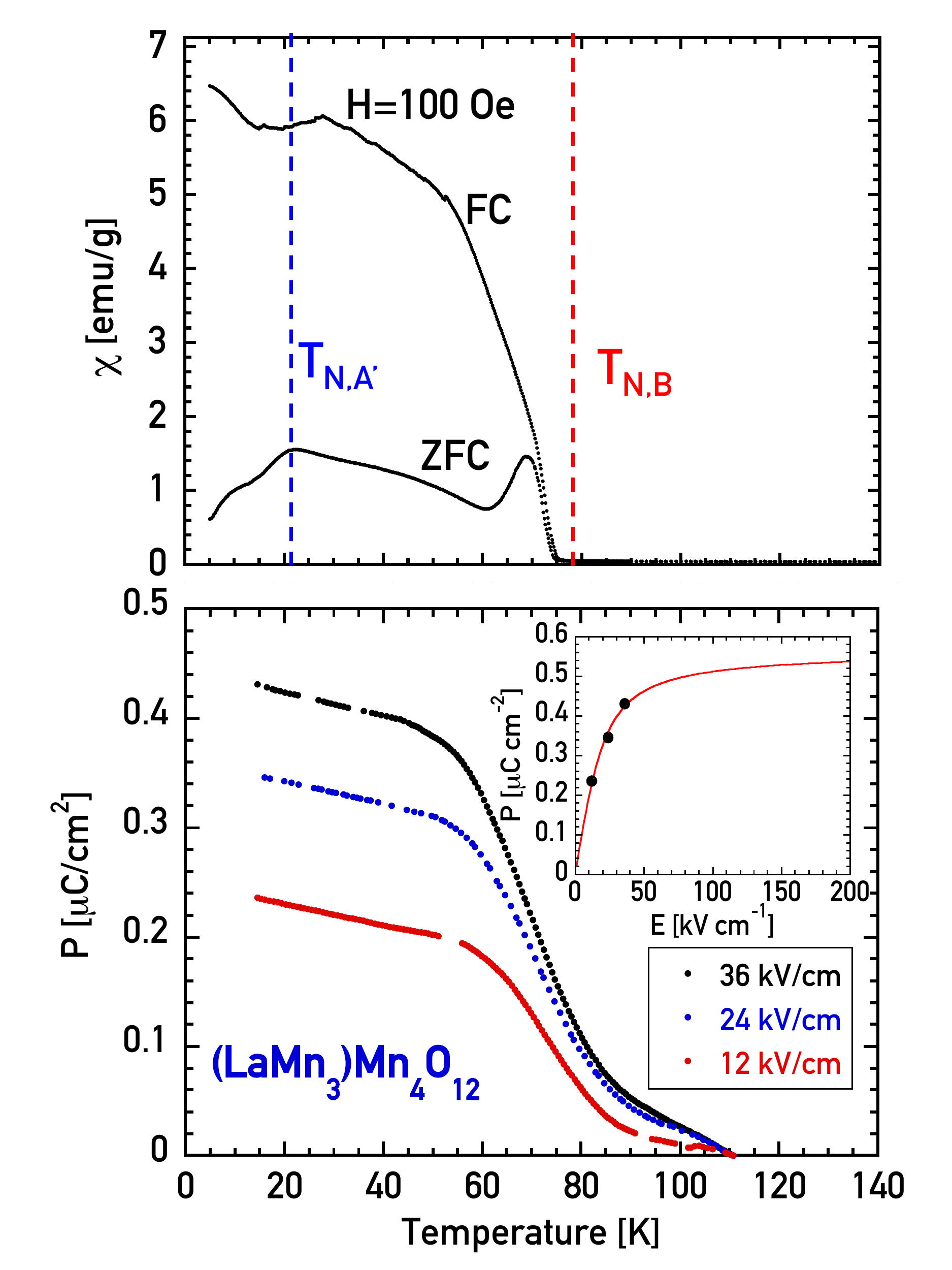}
\caption{\label{fig:pol} Top panel: temperature dependence of the magnetic susceptibility for a representative LMO single crystal measured at a 100 Oe field. Vertical broken lines indicate the two AFM ordering temperatures, $T_{N,A'}$ and $T_{N,B}$ of the Mn$^{3+}$ ions in the $A'$- and $B$-sites, respectively. Bottom panel: remnant polarization, $P$, of samples nr. 1 for different poling fields of 12, 24, 36 kV cm$^{-1}$ and for a heating rate of 0.4 K/min. The curves are obtained by integrating the pyrocurrent curves. Inset: dependence of $P$ upon poling field. A fit of the experimental points by using a Langevin function (solid line) yields a saturation value $P = 0.56$ $\mu$C cm$^{-2}$.}
\end{figure}

In Fig. \ref{fig:cellpar}, we show the refined unit cell parameters as a function of temperature in the 9-300 K range. Within the experimental resolution, no anomalous behavior is found for any of these parameters, except for the monoclinic angle, $\beta$, which displays an abrupt increase of slope at $T_{N,B}$, consistent with previous powder neutron diffraction data \cite{pro09}. This indicates a sizable magnetoelastic coupling which tends to stretch the $ac$-plane along the diagonal without appreciable change of unit-cell volume. Pyroelectric current, $I_p$, measurements show that this structural distortion is concomitant to the appearance of ferroelectricity, as shown in Figs. \ref{fig:pyro} and \ref{fig:pol}. Fig. \ref{fig:pyro} shows the temperature dependence of $I_p$ measured at different heating rates for both positive and negative poling fields of 1 kV mm$^{-1}$. The data consistently show a pronounced peak of $I_p$ in the vicinity of $T_{N,B}$, symmetric with respect to the polarity of the poling field, which gives evidence of ferroelectricity induced by the AFM ordering of the $B$-site Mn$^{3+}$ ions. The measurement yielded reproducible results for the two different samples nr. 1 a,d nr. 2 and was extended to high temperatures in order to rule out the possible contribution of leakage currents or space charges. For sufficiently low heating rates, the pyroelectric current vanishes above 100 K, which confirms the ferroelectric origin of the current. A possible contribution of thermally activated space charges to the pyroelectric current occurs only above 120 K. The peak displays a small shift towards high temperatures with increasing heating rate, which is due to a delay in the sample thermalization upon heating.       
            
In Fig. \ref{fig:pol}, we show the temperature dependence of the remnant polarization $P(T)$ obtained by integrating the pyroelectric current curve measured on sample nr. 1 at a heating rate of 1 K min$^{-1}$ for three different poling fields of 12, 24 and 36 kV cm$^{-1}$. As expected, the value of remnant polarization increases with field and a value as high as $P = 0.45$ $\mu$C cm$^{-2}$ is reached at the lowest temperature measured of 15 K. The dependence of $P$ on poling field plotted in the inset shows that the saturation of the polarization has not reached yet. A simple fit of this dependence using a Langevin function suggests a saturation polarization as high as $P = 0.56$ $\mu$C cm$^{-2}$. To the best of our knowledge, these are record value for magnetic ferroelectrics, twice as large as the value previously reported on CMO \cite{zha11,joh12}. Furthermore, as said before, about 6-10 times larger polarizations are expected in high-quality single crystalline film or bulk samples \cite{ish10}. Hence, in such samples of LMO, polarizations in the 1-10 $\mu$C cm$^{-2}$ range are expected.

In order to unveil the origin of the magnetic ferroelectricity observed, we investigated the existence of a magnetoelastic coupling by measuring the effect of magnetic field, $H$, on the low-temperature specific heat, $C(T,H)$. The motivation of these measurements is that this coupling is expected to alter the phonon frequencies and thus the lattice contribution to the specific heat. This qualitative picture is confirmed by Fig. 4, where we report the field dependence of $C(T,H)$ measured on the same batch of LMO powder samples at constant temperature in field-cooling mode. Note a dramatic reduction of $C$ with field, the relative reduction at 2 K being as large as 40 \% for a 9 T field. This reduction progressively decreases with increasing temperature below 20 K. This is due to the fact that the contribution of the AFM transition $T_{N,A'}$=21 K to the specific heat dominates over the magnetoelastic contribution.  

\begin{figure}[h]
\centering
\includegraphics[width=\columnwidth]{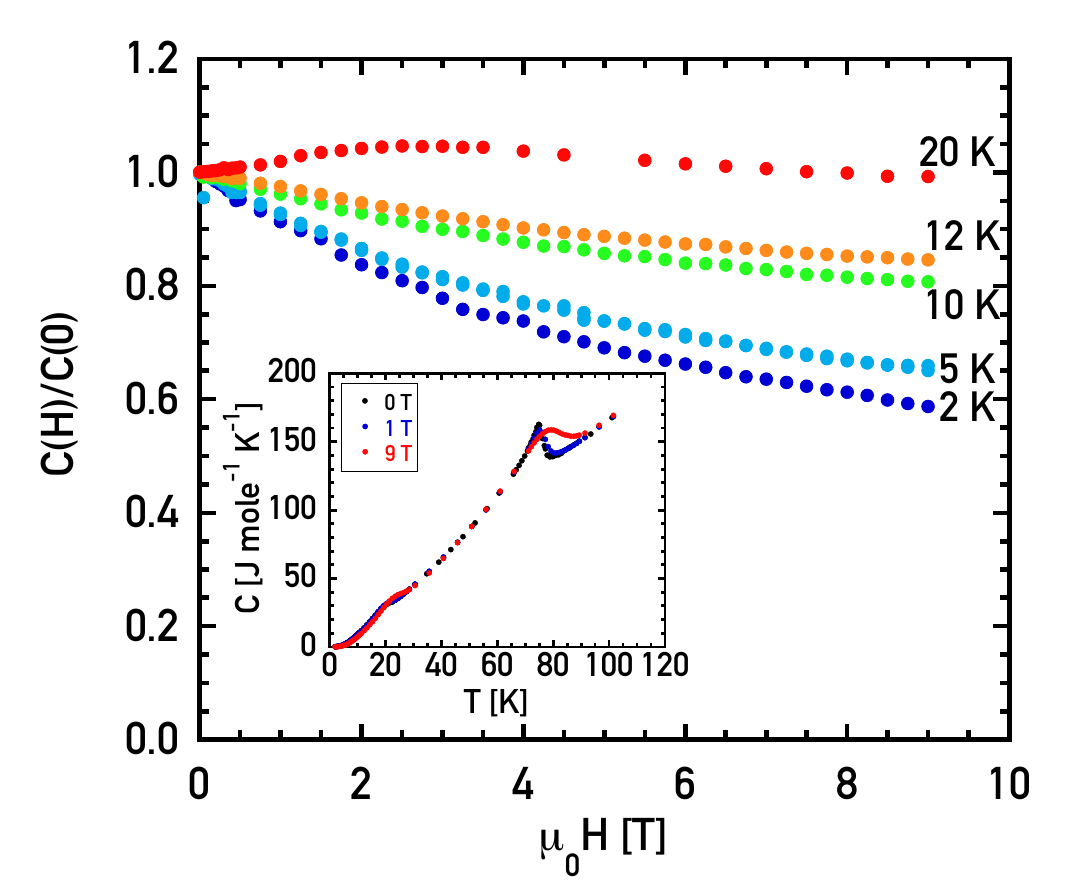}
\caption{\label{fig:Cp} Main panel: field dependence of the constant-pressure specific heat $C(T,H)$ at various temperatures. Note the large suppression of $C$ at low temperatures, where the difference between $C$ and the constant-volume specific heat can be neglected. Inset: temperature dependence of $C$ at different fields. Note the jumps at the two AFM ordering temperatures, $T_{N,A'}$ and $T_{N,B}$ of the Mn$^{3+}$ ions in the $A'$- and $B$-sites, respectively, smeared by the field.}
\end{figure}

We explain the above results by Callen and Callen's theory \cite{cal63} of magnetoelastic coupling developed for cubic crystals, suitable for the present pseudocubic LMO system. The theory describes the magnetoelastic (ME) energy as an expansion of orthogonal modes for the strain field:

\begin{equation}
H_{ME}=-\sum_{\mu}\sum_{j,l}B_{j,l}^{\mu} \sum_i \varepsilon_{i}^{\mu,j} \mathfrak{K}_i^{\mu,l}  
\end{equation}

where $B_{j,l}^{\mu}(T)$ are temperature-dependent magnetoelastic coefficients, $j$ numbers the modes of a given irreducible representation $\Gamma_{\mu}$ (IR) of the strain coordinates $\varepsilon_{i}^{\mu,j}$, $i$ numbers the dimensions of the given IR, $\mathfrak{K}_i^{\mu,l}$ are tensor cubic operators (TKOs) of degree $l$ forming an orthogonal basis. The Hamiltonian contains only even-degree TKOs, since the Hamiltonian is time-invariant. At low temperatures, the theory predicts that the $B_{j,l}^{\mu}(T)$ coefficients scale with the magnetization of the sublattice, consistent with the $l(l+1)/2$ power law originally established by Kittel and van Vleck \cite{kit60}:

\begin{equation}
\frac{B_{j,l}^{\mu}(T)}{\bar B_{j,l}^{\mu}(0)} = \left [ \frac{M(T)}{M(0)} \right ]^{l(l+1)/2}
\end{equation}

By knowing the elastic constants of the compound, the above expressions (1-3) enable to calculate all thermodynamic quantities like the specific heat of interest in the present work. Since these constants are not available for LMO, we limit ourselves to analyze the power-law behavior of the $C(H)$ curve expected from the above theory. To do so, we consider the external strain caused by an external magnetic field, $H$, which corresponds to the $j=0$ strain coordinates for a given IR. The leading contribution is given by the lowest-order ($l$=2) mode present in the quadrupolar ($E_g$) $\Gamma_{\gamma}$ IR, which is expected to give a characteristic $\sim M(T)^3$ dependence \cite{cal63}. Since the free energy depends quadratically on the ME coefficients, the corresponding ME contribution to the specific heat is described by a sum of two power-laws:

\begin{equation}
\frac{C_{ME}}{T} \sim -\frac{\partial^2 \left ( \bar B_{0,2}^{\gamma} \right )^2 }{\partial T^2} \sim \left [ M^5 \frac{\partial^2M}{\partial T^2} + M^4 \left ( \frac{\partial M}{\partial T} \right )^2 \right ]
\end{equation}

The above expression indicates that, at a given temperature, the ME contribution $C_{ME}$ grows very rapidly with $M$ according to a power law of index either 5 or 6, depending on the precise temperature dependent of the magnetization curve $M(T)$. We apply these considerations to LMO, where magnetization measurements as a function of field show a characteristic kink-like dependence above a threshold field $H^{*} \approx 1$ T (see top panel of Fig. 4 and \cite{pro09}). At 5 K, the kink corresponds to a magnetization value of about 55 emu/g. At higher field, the magnetization increases linearly with field with no sign of saturation. This dependence is characteristic of a canted AFM, consistent with the canted AFM C-type structure of the Mn$^{3+}$ $B$ sites, as reported previously \cite{pro09}. Specifically, above $H^{*}$, the canted component of the magnetization is parallel to the field and the response becomes AFM-like. The above $M(H)$ dependence enables us to plot the specific heat $C(H)$ as a function of $M$ shown in the bottom panel of Fig. 5.

\begin{figure}[h]
\centering
\includegraphics[width=\columnwidth]{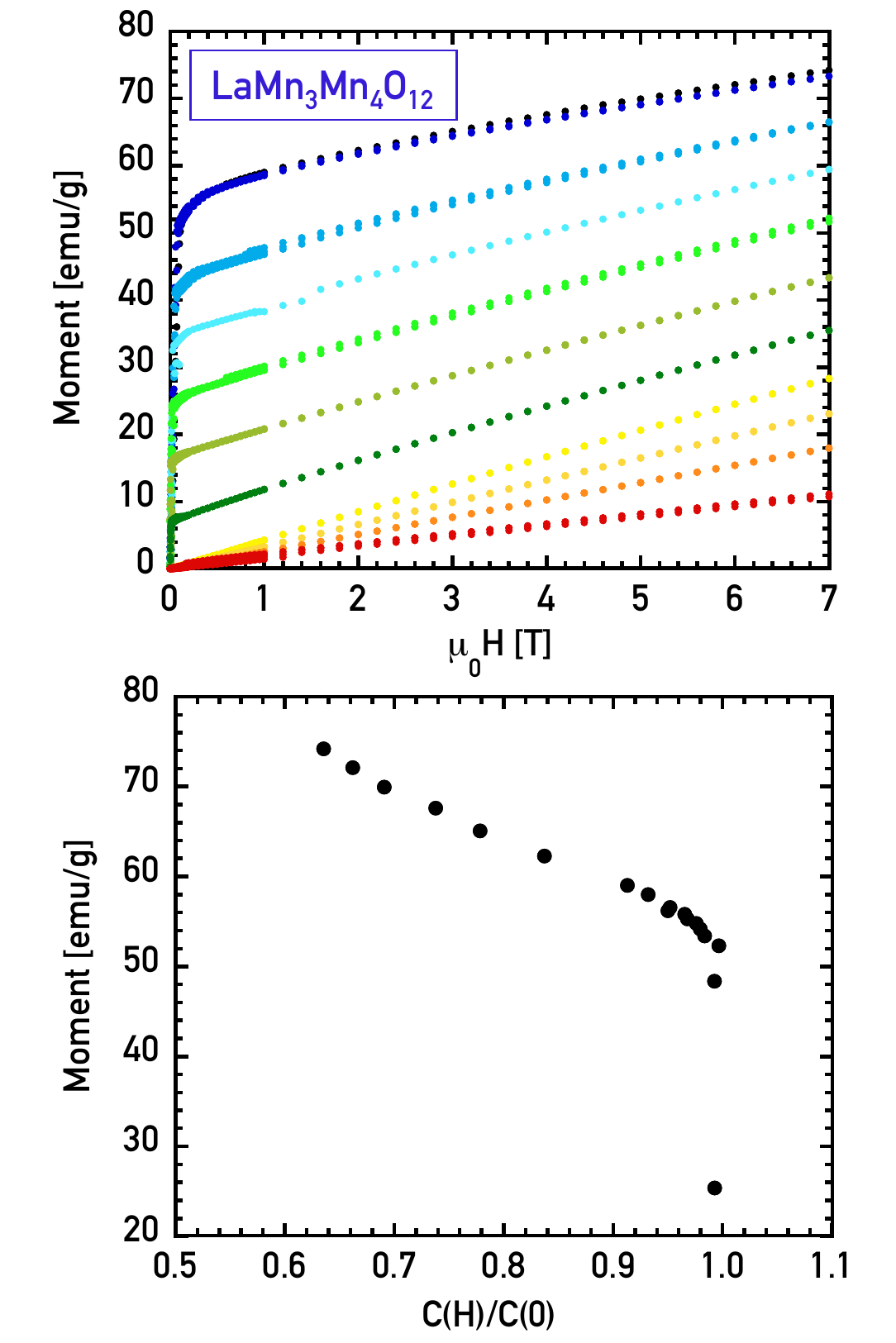}
\caption{\label{fig:MxH} Top panel: field dependence of the magnetization, $M$, of a representative LMO single crystal measured at constant temperature in the 5-300 K range. Bottom: specific heat measured at 5 K by varying field $H$ as a function of $M$. The curve is obtained by using the $M$ vs $H$ curve of the top panel.}
\end{figure}

Note the threshold behavior at a $M$ value of 55 emu/g, which coincides exactly with the kink of the magnetization curve of the top panel, followed by a very rapid decrease of the specific heat, consistent with the above prediction of high-order polynomial dependence. This behavior clearly indicates that the AFM-like magnetic response drives the reduction of the specific heat with field. A quantitative analysis of this dependence requires a measurement of the elastic constants, which goes beyond the scope of the present work.              

The observation of ferroelectric polarization induced by a canted AFM structure and the evidence of a large magnetoelectric coupling in LMO raise a number of questions.

(i) Considering that the AFM structure consists in the ordering of the $B$-site Mn$^{3+}$ ions and that this ordering is concomitant to a large magnetoelastic response of the monoclinic $\beta$ angle, the ferroelectric polarization should lie along the $ac$-diagonal. In the $ac$-plane, the magnetic structure displays AFM coupled collinear moments, suggesting that the macroscopic mechanism of spin-driven ferroelectricity is the symmetric exchange coupling, consistent with the large polarization observed. A contribution of the antisymmetric exchange (or inverse Dzyaloshinskii-Moriya) coupling, which typically leads to much smaller polarizations, should also be present because the ordering of the $B$-site Mn$^{3+}$ ions display a sizable canting along the $b$-axis direction \cite{pro09}.

(ii) The comparatively simple AFM structures of both $A'$- and $B$-ions offer ideal conditions to explain theoretically why only the latter structure induces ferroelectricity.       

(iii) The existence of a large polarization below $T_{N,B}$ implies a sizable noncentrosymmetric structural distortion below this temperature. It is known that it is experimentally difficult to detect these distortions by means of ordinary diffraction experiments and second-harmonic generation (SHG) experiments would be a more direct probe. In the present case, the SHG technique is not suitable, for LMO is a narrow-gap semiconductor with $\Delta$=60 meV, as indicated by resistivity measurements \cite{pro09}. A recent x-ray diffraction and optical conductivity study on single crystals \cite{ver18} indicates that the centrosymmetric $I2/m$ symmetry describes well the crystal structure in both, para- and ferro-electric phases above and below $T_{N,B}$. This conclusion is consistent with previous reports on the missing observation of noncentrosymmetric distortions in magnetic ferroelectrics. Specifically, only minute fm-size modulations of atomic displacements have been detected in the simple perovskite TbMnO$_3$ by means of resonant magnetic x-ray scattering \cite{wal11}. This discrepancy raises the question of the possibility of formation of polar domains not detected by diffraction or spectroscopic methods. 

(iv) The very large polarizations found in CMO and LMO suggest that manganese oxides with quadruple perovskite structure, ($A$Mn$_3$)Mn$_4$O$_{12}$, where $A$ is a di- or three-valent cation, are particularly favorable for hosting magnetic ferroelectricity. Further studies should take into account the peculiar features of quadruple perovskites not found in simple perovskites that may be associated with the above enhancement of magnetic ferroelectricity \cite{gau13}. 1) In all ($A$Mn$_3$)Mn$_4$O$_{12}$ compounds hitherto reported, no indication of oxygen defects has been found, which is attributed to the fact that oxygen vacancies would destabilize the bonding of the low-coordination number of the $A'$ sites. 2) In quadruple perovskites, the symmetry of the oxygen sites is higher than that of distorted simple perovskites, which limits strain-induced lattice distortions of the oxygen sublattice. As a result, the ability of the oxygen atoms to screen the electric dipole moment formed by the distortion of the Mn sublattice is reduced. An extensive low-temperature structural study of the ferroelectric phase of \lamno~ is required to verify this scenario.             

In conclusion, we have reported on ferroelectricity with a record value of remnant polarization, $P$=0.56 $\mu$C cm$^{-2}$ in the quadruple perovskite compound \lamno, concomitant to a canted $C$-type AFM ordering of the $B$-site Mn$^{3+}$ ions at $T_{N,B}$=78 K. This observation is accompanied by the evidence of a sizable magnetostriction of the unit cell and of a very large magnetoelastic coupling probed by field-dependent specific-heat measurements. This result is striking in two aspects. (i) Considering that the above polarizations, twice as large as that found in \camno~ single crystals, have been obtained on polycrystaline samples, 6-10 times larger values are expected in high-quality single crystalline film or bulk samples \cite{ish10}. This would correspond to extremely large values, $\sim 3-6$ $\mu$C cm$^{-2}$, comparable to those obtained in proper ferroelectrics, are expected, which would open the avenue towards multiferroic applications. (ii) If the above picture of polarization in the $ac$-plane will be confirmed, the collinear properties of the AFM order in the $ac$-plane driving the ferroelectricity in \lamno~ would rule out the scenario of inverse Dzyaloshinskii-Moriya effect and support the mechanism of direct exchange coupling between neighboring Mn$^{3+}$ ions. This result provides clear hints for the optimization of the properties of magnetic ferrolectrics and we envisage that even better multiferroic performances may be obtained in other quadruple perovskite compounds.
   
\bibliography{lamno}

\end{document}